\documentclass[12pt]{book}

\usepackage[dvips]{graphicx,color}
\usepackage{makeidx,observe,cosmology}

\makeauthorindex
\makeindex

\BookTitle{New Trends in Theoretical and Observational Cosmology}
\CopyRight{\copyright 2001 by Universal Academy Press, Inc.}

\begin{document}

\BookTitle{\itshape New Trends in Theoretical and Observational Cosmology}
\CopyRight{\copyright 2001 by Universal Academy Press, Inc.}
%\tableofcontents
\pagenumbering{arabic}

\chapter{H$_2$, D/H and the CMBR Temperature at\\ 
$z = 3.025$ Toward QSO 0347--3819}

\author{Sergei A. Levshakov\\
{\it National Astronomical Observatory, Mitaka, Tokyo 181-8588, Japan
and Physico-Technical Institute, St. Petersburg 194021, Russia}
\\
Paolo Molaro\\
{\it Osservatorio Astronomico di Trieste, Via G.B. Tiepolo 11, I-34131,
Trieste, Italy}
\\
Miroslava Dessauges-Zavadsky\\
{\it European Southern Observatory, Karl-Schwarzschild-Str. 2, D-85748,
Garching, Gernany and Observatoire de Gen\`eve, CH-1290 Sauverny, 
Switzerland}
\\
Sandro D'Odorico\\
{\it European Southern Observatory, Karl-Schwarzschild-Str. 2, D-85748,
Garching, Gernany} }
%
% Please note:
% One \AuthorContents{} is necessary
% for EACH CONTRIBUTION, for the contents page and
% One \AuthorIndex{} is necessary
% for EACH AUTHOR, for the index.
%
\AuthorContents{S. A.\ Levshakov, P.\ Molaro,
M.\ Dessauges-Zavadsky and S.\ D'Odorico} 

\AuthorIndex{Levshakov}{S.A.}
\AuthorIndex{Molaro}{P.} 
\AuthorIndex{Dessauges-Zavadsky}{M.} 
\AuthorIndex{D'Odorico}{S.}

\section*{Abstract}

A new molecular hydrogen cloud has been
identified at $z = 3.025$ in the absorption-line 
spectrum of the quasar 0347--3819 
observed with the UVES spectrograph at the VLT/Kueyen telescope.
At the same redshift numerous metal lines and the D\,{\sc i} Ly-5, Ly-8,
Ly-10, and Ly-12 lines were detected. 
The simultaneous analysis of metal and hydrogen lines yielded
D/H = $(3.75\pm0.25)\times10^{-5}$. 
This value is consistent with SBBN if the baryon-to-photon ratio,
$\eta$, lies within the range
$4.4\times10^{-10} \leq \eta \leq 5.3\times10^{-10}$, implying
$\Omega_b\,h^2_{100} = 0.018\pm0.002$ (1 $\sigma$ c.l.). 
The population of the ground state rotational levels of H$_2$ from
$J=0$ to 5 revealed a Galactic-type 
UV radiation field in the H$_2$-bearing cloud 
ruling out UV pumping as an important mechanism for C$^+$.
Accounting for two other important excitation mechanisms such as
collisions and FIR photon absorption, we were able
to measure
for the first time the temperature of the cosmic background radiation
at $z > 3$, $T_{\rm CMBR} = 12.1^{+1.7}_{-3.2}$~K, 
from the analysis of the C$^+$ fine-structure lines. 
This result supports the value of $T_{\rm CMBR} = 10.968\pm0.004$~K 
predicted at $z = 3.025$ by the hot Big Bang cosmology.

\section{Introduction}

In the standard Big Bang model (SBB)
the temperature of the relic radiation  from
the hot phase of the Universe is predicted to
increase linearly with redshift $z$:
$T_{\rm CMBR}(z) = T_{\rm CMBR}(0)\,(1+z)$.
At the present epoch direct measurements show that
$T_{\rm CMBR}(0) = 2.725\pm0.001$~K (1 $\sigma$ c.l.), 
and that the relic radiation follows
a Planck spectrum with very high precision \cite{Mather}.
However,
at earlier cosmological epochs $T_{\rm CMBR}$ cannot be
measured directly.
It was suggested that this scaling in proportion to 
$(1+z)$ can be tested by observing the population of
excited fine-structure lines in the QSO absorption spectra
\cite{BW}. 

The relative population of the fine-structure levels 
may not, however, be caused by photo-absorption of the
CMBR only.
Non-cosmological sources
such as particle collisions, pumping by UV radiation or by
IR dust emission  
may compete with the CMBR to populate the excited
fine-structure levels.
Only independent knowledge of the ambient radiation field, particle
densities and the kinetic temperature of the gas
allow to disentangle the  contribution of the background radiation
from that of other mechanisms.
For these reasons  previous studies set only
upper limits to $T_{\rm CMBR}$ 
\cite{Meyer}, \cite{Songaila},
\cite{Lu}, \cite{Ge}, \cite{Roth}, \cite{Ge2}.

The physical parameters in question can be
accurately estimated if the absorber is a diffuse molecular cloud.
Molecules allow to measure
the volumetric gas density $n_{\rm H}$, the
kinetic temperature $T_{\rm kin}$ and the intensity of the UV
radiation field through the analysis of their distribution on 
the low rotational levels. In particular, intervening
molecular clouds showing H$_2$ and   
the fine-structure lines of
C\,{\sc i} and C\,{\sc ii} provides a unique opportunity to
measure the cosmic microwave background radiation temperature in early
cosmological epochs and to test the SBB predictions.

Another observational test of the SBB model is the measurement
of the hydrogen isotopic ratio at high $z$.
The standard (homogeneous) Big Bang nucleosynthesis (SBBN) predicts
the same D/H abundance ratio for any direction in the early Universe
since ``no realistic astrophysical process other than the Big Bang could
produce significant D'' \cite{Schramm}.
Deuterium is created exclusively in BBN and therefore
we can expect that the D/H ratio decreases with cosmic time 
due to conversion
of D into $^3$He and heavier elements in stars.
It is clear that the precise measurements
of the D/H values at high redshift are extremely 
important to probe whether BBN
was homogeneous.
The choice of the appropriate BBN model
may in turn place constrains on different
models of structure formation.

In this contribution we discuss the role of the H$_2$-bearing cloud
with respect to the measurement of D/H and $T_{\rm CMBR}$
at $z_{\rm abs} = 3.025$ toward the quasar 0347--3819.

\section{Observations and results}

The spectroscopic observations of Q0347--3819
($z_{\rm em} = 3.23$, V = 17.3)
obtained with the VLT 8.2~m telescope
are described in \cite{DDM} and \cite{lev02}.
The spectrum was obtained with the
resolutions FWHM $\simeq 7.0$ km~s$^{-1}$ and $\simeq 5.7$ km~s$^{-1}$
in the UV and near-IR ranges, respectively.
A signal-to-noise
ratio of S/N $\simeq 10-40$ per resolution element
was achieved in these observations.

The analysis of the H$_2$, deuterium and metal absorption-lines
was performed in \cite{lev02}. We found that 
the fractional abundance of H$_2$ 
is $f_{{\rm H}_2} = (3.3\pm0.2)\times10^{-6}$. 
This value is similar to that 
observed in the Galactic diffuse clouds with low color
excesses, $E(B-V) < 0.1$. 
The kinetic gas temperature is found to be less than 430~K.
The population of the
low rotational levels is represented by
a single excitation temperature of $T_{\rm ex} =
825\pm110$~K. 
The observed
population ratios of H$_2$ in different rotational states and
the C\,{\sc ii}$^\ast$/C\,{\sc ii} ratio give the
number density 
$n_{\rm H} \simeq 6$~cm$^{-3}$,
and the cloud dimension along the line of sight
$D \simeq 14$ pc. 
The relative populations of H$_2$ in the $J=4$ and 5
rotational levels correspond to
the rate of
photo-destruction $I \simeq 2\times10^{-10}$ s$^{-1}$.
The photo-absorption rate, $\beta_0 \simeq I/0.11$, is thus
equal to $\beta_0 \simeq 2\times10^{-9}$ s$^{-1}$.
Taking into account that the Galactic interstellar radiation field value
ranges in
$5\times10^{-10}$ s$^{-1}$ $\leq \beta_0 \leq 1.6\times10^{-8}$ s$^{-1}$
\cite{Jura},
we conclude that the UV radiation fields in the
$z_{\rm abs} = 3.025$ absorbing cloud and
in the Galactic ISM are very much alike.
We also found that the formation rate of H$_2$
upon grain surfaces, $R\,n_{\rm H} \simeq 3\times10^{-16}$ s$^{-1}$,
is similar to that measured in the Galaxy \cite{Jura}.

The simultaneous analysis of metal profiles and the
H+D Lyman series lines yielded a new estimation of the
hydrogen isotopic ratio
D/H = $(3.75\pm0.25)\times10^{-5}$ in this DLA.
It implies that 
the present-day baryon density is
$\Omega_{\rm b}\,h^2 = 0.018 \pm 0.002$.
The obtained D/H value is in good agreement with the
hydrogen isotopic ratio found in the Lyman Limit Systems in
\cite{Tytler}, \cite{lev98}, \cite{lev00}.
We thus summarize that when all most accurate measurements of D/H in high
redshift QSO absorbers are considered, one finds the {\it same} D/H 
which is about 2.5 times the mean ISM value of
D/H = $(1.50\pm0.10)\times10^{-5}$ \cite{Linsky}.

Recently, an attempt was made to measure $T_{\rm CMBR}$ in the 
H$_2$ molecular cloud at
$z = 2.34$ toward Q1232+0815 \cite{SPL}.
However, further analysis of this system \cite{var} revealed
inconsistency of the H$_2$ and HD column density measurements 
\cite{mldd} which places the value of $T_{\rm CMBR}$
found in \cite{SPL} to an upper limit only.

In our analysis of the DLA system at $z = 3.025$ \cite{mldd}
we obtained a self-consistent solution for both molecular and
metal absorption lines. We have considered collisions and excluded
fluorescence and dust emission as significant processes in the
population of the C$^+$ excited level $J=3/2$. 
Our measurement of
$N$(C\,{\sc ii}$^\ast$)/$N$(C\,{\sc ii})
= $(3.8\pm0.3)\times10^{-3}$ leads to the
most probable value of  
$T_{\rm CMBR} = 12.1^{+1.7}_{-3.2}$~K
which is fully consistent with the
predicted temperature $T_{\rm CMBR} = 10.968\pm0.004$~K
within 1 $\sigma$ confidence intervals.

We conclude that 
our result, together with upper limits measured in the above 
mentioned papers
support the linear evolution of the CMBR
within the framework of the hot Big Bang cosmology.
Being considered together, these D/H and $T_{\rm CMBR}$ measurements
show again that the fundamental concepts of cosmology and of
Big Bang nucleosynthesis are consistent with observations.

\end{document}